# On the reactivity of low coordinated atoms on foreign solid substrates as models of single atom catalysts

Ana S. Dobrota[1], Igor A. Pašti[1,2]*, Aleksandar Z. Jovanovi [1], B rje Johansson[2,3], Natalia V. Skorodumova[2,3]

[1]*University of Belgrade – Faculty of Physical Chemistry, Belgrade, Serbia*

[2]*Department of Materials Science and Engineering, School of Industrial Engineering and Management, KTH - Royal Institute of Technology, Stockholm, Sweden*

[3]*Department of Physics and Astronomy, Uppsala University, Uppsala, Sweden*

**Abstract**

Catalysis has entered everyday life through a number of technological processes relying on different catalytic systems. The increasing demand for such systems requires rationalization of the use of their expensive components, like noble metal catalysts. As such, a catalyst with low noble metal concentration, in which each one of the noble atoms is active, would reach the lowest price possible. Nevertheless, there are no reactivity descriptors outlined for this type of low coordinated supported atoms. Using DFT calculations, we consider three diverse systems as models of single atom catalysts. We investigate monomers and bimetallic dimers of Ru, Rh, Pd, Ir and Pt on MgO(001), Cu adatom on thin Mo(001)-supported films (NaF, MgO and ScN) and single Pt adatoms on oxidized graphene surfaces. Reactivity of these metal atoms was probed by CO. In each case we see the interaction through the donation-backdonation mechanism. In some cases the CO adsorption energies can be linked to the position of the d-band center and the charge of the adatom. Higher positioned d-band center and less charged supported single atoms bind CO weaker. Also, in some cases metal atoms less strongly bonded to the substrate bind CO more strongly. The results suggest that the identification of common activity descriptor(s) for single metal atoms on foreign supports is a difficult task with no unique solution. However, it is also suggested that the stability of adatoms and strong anchoring to the support are prerequisites for the application of descriptor-based search for novel single atom catalysts.

**Keywords:** supported single atoms; single atom catalysts; reactivity descriptor; d-band center; adatom charge

---

* **corresponding author:** e-mail: igor@ffh.bg.ac.rs

## 1. Introduction

Many important technological processes depend on the use of expensive catalytic components like noble metals, inducing both practical and economical limitations to their wide use. For these reasons the reduction of expensive catalytic components' content is a motivation for intensive scientific research, bringing experts from different fields to the same table.

In the case of heterogeneous catalysis, the catalyst activity is governed by its surface. Exposing more of the surface to the reactants increases the number of active sites and improves catalytic activity. This is achieved by dispersing catalyst particles over a suitable support. Until particle sizes are large enough, their behavior can be understood based on the properties of the bulk catalyst, taking into account the number of exposed active sites.[1] However, once the dispersion is sufficiently high, to understand catalytic performance[1] other factors must be taken into account, including the specific atomic configuration, the number of atoms in the particle/metal cluster,[2] and their interactions with the support.[3,4] Such a situation is driven to the extreme when the catalyst is found in the form of single supported atoms on a foreign substrate.[1,5] While single-atom catalysts (SAC) have already been found in practical applications[6,7] recent advancements produced a great deal of new SACs and catalytic systems in which these are applied.[5] A comprehensive overview of the most recent results can be found in the papers by Liu[1], Gates *et al.*[5] and Li *et al.*[8]

Considering monoatomic dispersion, SAC maximize the utilization of (expensive) metal components in the catalyst, but also provide the strategy to tune the activity and selectivity of a catalytic reaction[1]. One of the main challenges for the practical use of such catalysts is the necessity of strong anchoring of the single metal atoms to the support, followed by keeping them stable and functional under reaction conditions.[1,9] In addition to these issues, Gates *et al.*[5] have listed a number of questions which must be considered when developing atomically dispersed supported metal catalysts. These can be classified in those related to (i) synthesis, (ii) characterization and identification of their properties, (iii) catalyst stability and (iv) integration into more complex systems. In our view, among other questions, the ones related to the identification of crucial properties of SACs and the confirmation that the atomically dispersed species are indeed catalytically active stood out as particularly important.

As in heterogeneous catalysis the interaction between the catalyst and reactants/intermediates plays a decisive role in catalytic performance, the reactivity of single supported atoms must be understood. There are numerous examples of successful use of highly dispersed metals and SACs for different catalytic reactions[10–13] but there are no general strategies to predict reactivity of supported metal atoms necessary for the rational design of such materials. Namely, in the case of extended solid catalysts numerous activity descriptors have been identified[14–20] offering clear links between the electronic structure, chemisorption properties and the reactivity. In contrast, in the case of single supported atoms there are no clear indications which parameters can be uniquely linked to their reactivity. Charge state of supported atoms was often outlined as an important parameter.[21–24] It was considered that extra charge can boost bond cleavage in molecular adsorbates[25] and alter chemisorption properties of supported metals.[26] Our work on MgO(001)-supported bimetallic dimers suggested that for such systems the d-band centers can be linked to reactivity of low coordinated atoms.[27] Recently, combined experimental and theoretical study by Vorobyeva *et al.*[28] pointed out the metal (Pd) oxidation state, assessed by XPS, as an important activity descriptor for the semi-hydrogenation of 2-methyl-3-butyn-2-ol.

In this contribution we consider three fundamentally different metal@support systems, analyze the properties of supported metal atoms and probe their reactivity. For each system we attempt to link the reactivity trends to bonding and the electronic structure parameters of supported metal atoms.

**2. Computational details**

Here we analyze the reactivity of three rather diverse low coordinated supported metal atom systems. First, we address the reactivity of single metal atoms and bimetallic dimers supported by MgO(001). Our previous combinatorial study of MgO(001)-supported metal dimers suggested that d-band centers of individual atoms can be used to predict their chemisorption properties.[27] We probe the reactivity using CO molecule. Next, we analyze the effects of Cu adatom charging on its chemisorption properties. Controllable charging of Cu is accomplished by changing the support. In specific we have used 2 monolayers (ML) thick overlayers (NaF, MgO and ScN) over Mo(001) substrate as described in ref.[29]. For these cases we use CO and atomic H to probe the reactivity of single Cu adatoms. Finally, we investigate the reactivity of single Pt atoms supported by oxidized graphene using CO as the probe adsorbate.

*2.1. MgO(001) supported metal atoms and bimetallic dimers*

The Density Functional Theory (DFT) calculations were performed within the generalized gradient approximation (GGA) employing Perdew–Burke–Ernzerhof (PBE) exchange correlation functional.[30] The calculations were done using the PWscf code of Quantum ESPRESSO distribution.[31,32] The kinetic energy cutoff for the selection of the plane-wave basis set was 28 Ry and the charge density cutoff was 16 times higher. The lattice constant of MgO was calculated to be 4.22 Å, in good agreement with the experimental value of 4.21 Å.[33] The MgO (001) surface was modeled as a three layer slab with the (2×2) unit cell. Monomers and dimers of investigated transition metals were placed on one side of MgO(001) in the geometry corresponding to that described in ref.[27]. Full relaxation was allowed for all the atoms except the ones in the bottom layer of MgO(001). The Brillouin zone was sampled using the (2×2×1) Monkhorst-Pack grid.[34] Vacuum thickness and the dipole correction were adjusted to prevent supercells' interaction along the z direction.[35] The model of the MgO(001) surface used here is described in detail in ref.[36]. Additional information regarding the accuracy of the model can be found in ref.[27].

*2.2. Cu adatom on Mo-supported thin films*

DFT calculations were performed using the projector augmented wave method[37] in combination with the GGA approximation in the PBE parametrization[30] as implemented in Vienna Ab initio Simulation Package.[38–42] The cutoff energy of 800 eV was used in the calculations. The Brillouin zone was sampled over the (8×8×1) Pack–Monkhorst k-points mesh.[34] We have used symmetric slab consisted of Mo(001) (7 layers) and two layers of NaF, MgO or ScN on each side of the Mo slab. Cu adatom was also placed on both sides of the slab. During geometry optimisation the relaxation of all atoms, except for the Mo atoms in the central Mo layer, was allowed. More details about this set of calculations can be found in ref.[29].

*2.3. Single Pt atoms on oxygen-functionalized graphene*

The DFT calculations were carried out within GGA-PBE approach[30] using ultrasoft pseudopotentials as implemented in the PWscf code of Quantum ESPRESSO distribution.[31,32] The Kohn-Sham orbitals were expanded in a plane wave basis set with the kinetic energy cutoff of 30 Ry, while the charge density cutoff was 448 Ry. Pristine graphene was modeled as 32 carbon atom layer within an orthorhombic 9.88×8.65×16 Å supercell, as used previously in refs.[43,44]. Oxidized graphene (*ox*-graphene) was modeled as *p*-graphene with epoxy or hydroxyl groups attached to both sides of the layer. The detailed description of the surface models is given in ref.[44].

*2.4. Interactions quantified*

In order to quantify the interaction of low coordinated atoms with the support we define metal binding energy ($E_b$(M)) as:

$$E_b(M) = E_{M@subs} - E_{subs} - E_M \qquad (1)$$

$E_{M@subs}$, $E_{subs}$ and $E_M$ stand for the total energy of metal atom M on foreign substrate, the total energy of substrate and the total energy of isolated atom M. Reactivity was probed using CO for all the studied metal atoms and using atomic H for Cu on layered substrates. Reactivity is quantified using the adsorption energy of adsorbate A (CO or H) as:

$$E_{ads}(A) = E_{A+M@subs} - E_{M@subs} - E_A \qquad (2)$$

$E_{A+M@subs}$ and $E_A$ are the total energy of metal atom M on foreign substrate with A adsorbed and the total energy of isolated adsorbate. In order to characterize the electronic structure of supported adatoms we calculated their d-band centers ($E_{d\text{-band}}$) as the first moments of the d-bands and the charge states. The charge transfer was analyzed using the Bader algorithm[45] on a charge density grid by Henkelman *et al.*[46] Graphical presentation was made using VMD code.[47]

## 3. Results and discussion
### 3.1. Reactivity of metal atoms and dimers on MgO(001)

In the first part we analyze a model case of oxide-supported metal catalysts. MgO(001) is chosen as support due to its importance in surface science and catalysis.[23,48–50] We analyzed the reactivity of supported single Ru, Rh, Pd, Ir and Pt atoms, metals which of great importance for many contemporary technologies. As we reported previously,[27] these metal adatom prefer adsorption on O-site of pristine MgO(001) surface and bond to it with a variable strength (Table 1).

**Table 1.** Binding energies of the selected single atoms on MgO(001), their charge states ($q$) and d-band centers, along with CO adsorption energies on M@MgO(001) and M-CO bond energies in the gas phase ($E_{M\text{-}CO}$).

|  | $E_b(M)$ / eV[a] | $q(M)$ / e[a] | $E_{d\text{-band}}(M)$ / eV[a] | $E_{ads}(CO)$ / eV[a] | $E_{M\text{-}CO}$ / eV |
|---|---|---|---|---|---|
| **Ru** | −1.09 | −0.28 | −0.80 | −3.09 | −2.77 |
| **Rh** | −1.68 | −0.28 | −0.83 | −3.12 | −2.97 |
| **Pd** | −1.35 | −0.24 | −1.49 | −2.49 | −2.24 |
| **Ir** | −1.98 | −0.46 | −0.99 | −3.79 | −3.48 |
| **Pt** | −2.27 | −0.45 | −1.44 | −3.63 | −3.42 |

[a] ref.[27]

Certain amount of charge is transferred to each metal adatom and it ranges from 0.24 |e| in the case of Pd to 0.46 |e| in the case of Ir adatom. We have calculated the d-band centers for the studied adatoms. The trends in the d-band center values do not follow the ones of pure metallic surfaces.[51] Next, we calculated the CO adsorption energies on supported metal atoms. A rather strong interaction with CO can be seen, which is much stronger than the interaction of CO with clean metallic surfaces.[51,52] This is understandable taking into consideration the low coordination of supported metal atoms compared to the atoms at metal surfaces. Nevertheless, we could not correlate CO adsorption energies on low coordinated atoms to those on the corresponding metallic surfaces. However, we have calculated the energy of the M-CO bond in the gas phase and found that CO adsorption energies on M@MgO(001) systems correlate well with the M-CO bond energy (Fig. 1).

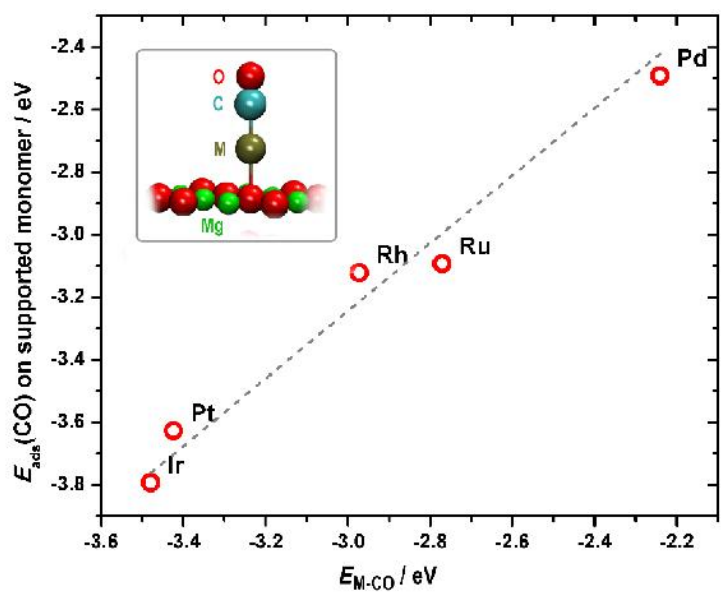

**Figure 1.** CO adsorption energies on supported metal monomers *vs.* M-CO binding energy in the gas phase. The inset shows an example structure of CO adsorbed at M@MgO(001) system.

While single supported atoms can bind CO with different strengths, the question is how to further modulate the interaction strength. In practice this would relate to fine tuning of reactivity and catalytic activity. For this purpose we turn to the case of MgO(001) supported bimetallic dimers (M1M2) and probe the reactivity of given atoms in dimers using CO. Dimers oriented parallelly to MgO(001) are considered. The probe molecule was placed on top of M1 (M1 = Ru, Rh, Pd, Ir, Pt) atom in M1M2 dimer (M2 = Ru, Rh, Pd) and the structure was fully relaxed. Here we do not investigate the ways CO can approach the dimers, or a possibility that CO can simultaneously attach to the dimer and the MgO support, although such bonding is important in heterogeneous catalysis.[53–55] By changing the chemical environment of the considered M1 atoms it is possible to adjust their charge state and d-band center positions (Table 2).[27] We have calculated the binding energy of each M1 to M2@MgO(001), where the final result is the considered M1M2@MgO(001) dimer (Table 2). All these properties show certain regularities. Namely, we see that M1 binding energy to M2@MgO(001) decreases as M2 goes from Ru to Pd, just as for the case of dimers in the gas phase.[27] In the same order the charge transferred to M1 upon its bonding to M2@MgO(001) decreases and the d-band center shifts to higher values.

Table 2. M1 binding energies to M2@MgO(001) complex, charge state and d-band centers of M1 atoms in M1M2@MgO(001) dimers and CO adsorption energies at M1 atoms of dimers.

| M1 | M2 | $E_b$(M1) / eV | $q$(M1) / e[b] | $E_{d\text{-band}}$(M1) / eV [b] | $E_{ads}$(CO) / eV |
|---|---|---|---|---|---|
| **Ru** | Ru | −4.12 | −0.23[a] | −1.93 | −3.24 |
|  | Rh | −3.38 | −0.08 | −1.66 | −3.31 |
|  | Pd | −2.11 | −0.12 | −1.42 | −3.47 |
| **Rh** | Ru | −4.15 | −0.43 | −1.83 | −3.00 |
|  | Rh | −3.43 | −0.28[a] | −1.55 | −3.11 |
|  | Pd | −2.56 | −0.23 | −0.91 | −3.26 |
| **Pd** | Ru | −2.71 | −0.37 | −1.90 | −1.93[b] |
|  | Rh | −2.39 | −0.30 | −1.58 | −2.14[b] |
|  | Pd | −1.83 | −0.23[a] | −1.37 | −2.46[b] |
| **Ir** | Ru | −5.01 | −0.59 | −2.29 | −3.42 |
|  | Rh | −4.11 | −0.53 | −1.99 | −3.74 |
|  | Pd | −3.05 | −0.42 | −1.36 | −3.92 |
| **Pt** | Ru | −4.15 | −0.66 | −2.23 | −2.64[b] |
|  | Rh | −3.62 | −0.57 | −1.92 | −3.02[b] |
|  | Pd | −2.84 | −0.55 | −1.70 | −3.35[b] |

[a]average for two atoms in homonuclear dimer; [b]ref.[27]

It can be observed that CO binding at M1 increases in strength as the M1 binding to M2@MgO(001) weakens. This is can be understood intuitively: more saturated M1 is less reactive. In addition, we also observe the link between $E_{ads}$(CO) and the charge and d-band center of M1. In particular, we observe that CO adsorption is stronger on M1 with less charge. In the same way, we observe the link between the CO adsorption energies and the d-band center of M1 (Fig. 2), which was previously described for CO adsorption on PtM2 and PdM2 dimers.[27]

The behavior is the same as in the case of metallic surfaces: as the d-band center shifts towards the Fermi level, the interaction with CO becomes stronger.[27,56,57]

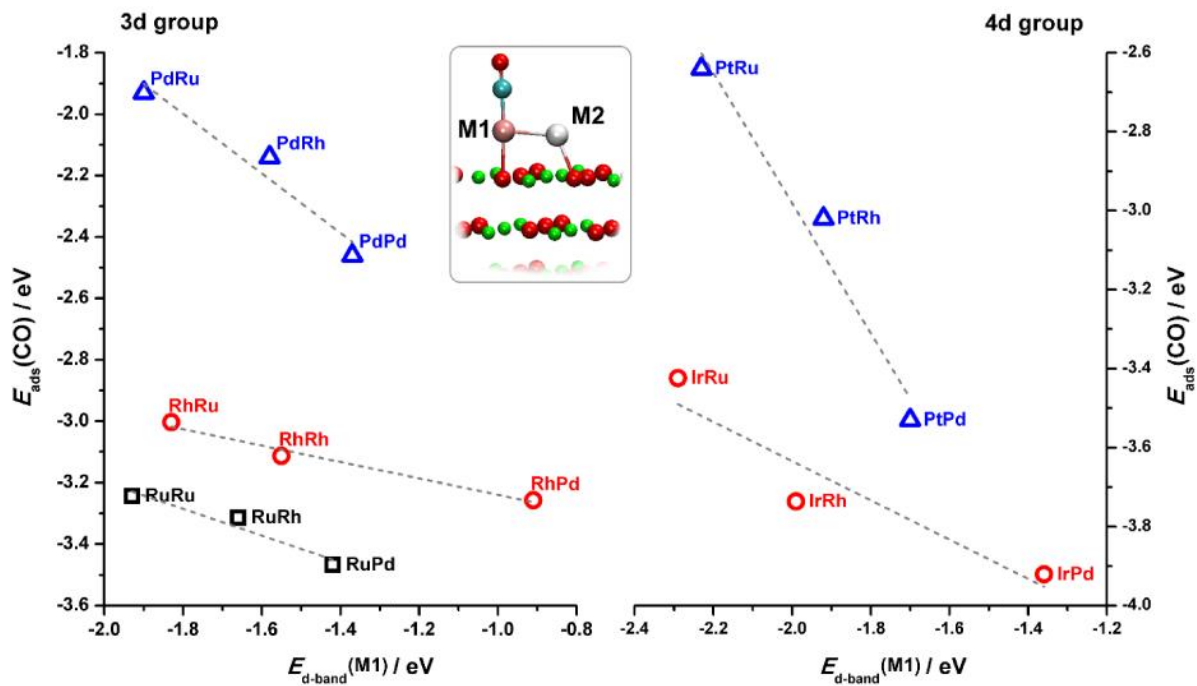

**Figure 2.** Correlation of the CO adsorption energy on the M1 atom of the M1M2 dimer supported on defect-free MgO(001) (inset) with the d-band center of M1.

In order to analyze the observed trends we turn to the investigation of the electronic structure of adsorbed CO and supported metals. Also, we analyze charge redistribution patterns upon the adsorption of CO (Fig. 3). We observe that in all the cases certain amount of charge is transferred to the adsorbed CO molecule, resulting in the elongation of the C–O bond (in all the cases bond length in adsorbed CO molecule is around 1.16 Å or slightly larger, while in isolated CO it is 1.134 Å[58]). This indicates that the CO chemisorption can be described within Blyholder model as coupling of the CO 5σ (donation) and 2π* states (backdonation) to the metal d-valence states[59] in the case of considered low coordinated atoms. Such behavior is clearly seen from Fig. 3, where the d-states of Ir atom (either single supported atom or in IrRh dimer) show strong coupling with the states of the adsorbed CO molecule. It is also obvious that the presence of Rh has a great impact on the electronic structure of Ir in the dimer, so that the d-band center is shifted away from the Fermi level significantly. However, the reactivity, as seen by $E_{ads}$(CO), is similar to that of isolated Ir atom on MgO(001) (Tables 1 and 2). Although the connection between the d-band center and the CO adsorption energies can be useful, obviously it cannot fully describe the behavior of low coordinated atoms, as monomers generally do not follow the trends given by the same atoms in dimers.

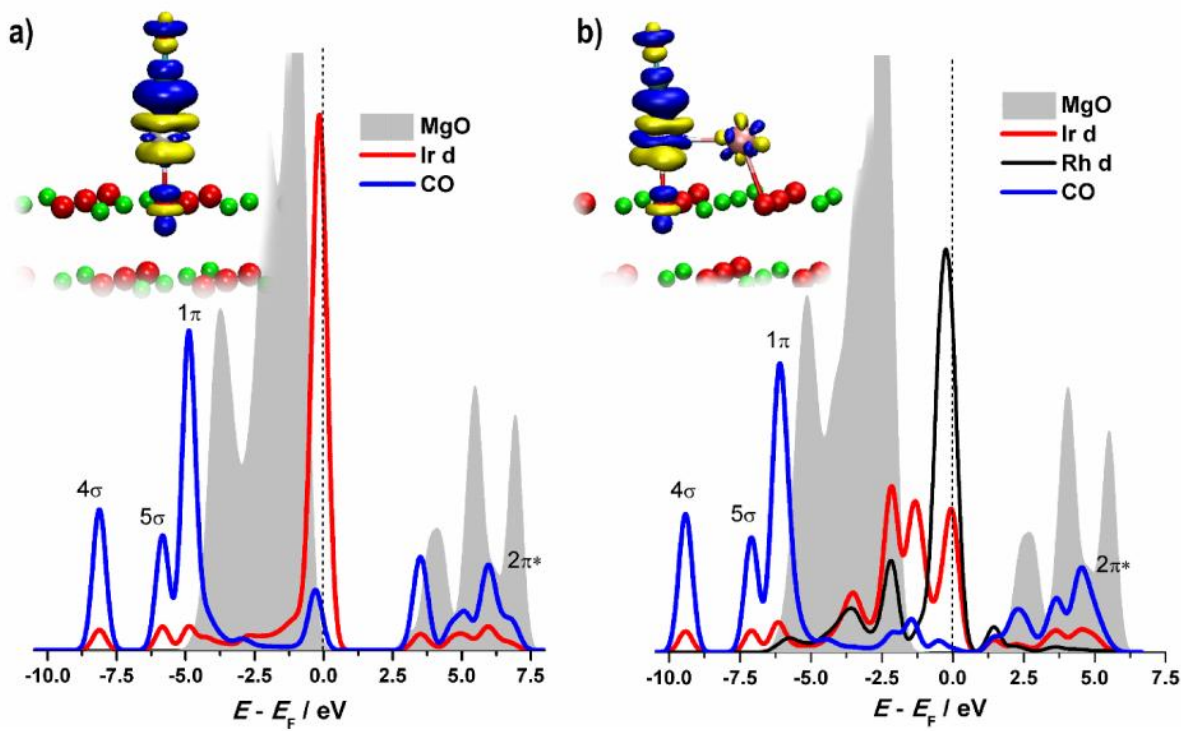

**Figure 3.** Projected density of states for CO adsorption on MgO supported Ir monomer (a) and IrRh dimer (b). Insets give charge redistribution patterns with blue colored regions indicating the increase of the charge density and yellow colored regions indicating depleted charge density (isosurface values are ±0.007 e Å$^{-3}$).

In contrast to d-band centers, we find that the charge state of considered atoms (Ru, Rh, Pd, Ir and Pt) can be generally linked to the adsorption energy of CO on a given atom. As a simple rule of thumb we see that more negatively charged atoms bind CO less strongly. Similar observations were previously obtained, for example, in the case of charged Au clusters in the gas phase.[60] Positively charged atoms and clusters were found to bind CO much stronger than negatively charged counterparts, suggesting that the forward donation from CO is the dominant mechanism of interaction between $Au_n^{\pm}$ and CO. This was also possible to see through the elongation of the C–O bond which was more prominent in the case of negatively charged $Au_n$ clusters.[60] Here we see the same behavior: in the case of Ir monomer on MgO(001) (charge –0.46e) C–O bond length is 1.17 Å, while in the case of Ir in Rh dimer (charge –0.53|e|) C–O bond length is 1.18 Å. The effects of charging on CO adsorption were also analyzed and

compared for the cases of Au and Pd surfaces where strong effect was seen for Au and small for Pd.[61]

**3.2. Tuning the reactivity by charging metal atoms - Cu adatom on Mo-supported NaF, MgO and ScN thin films**

In the next step we further explore the effect of charging supported single metals atoms on their chemisorption properties. We employ the phenomena of adatom charging when on thin insulating metal-supported films. We do such analysis for Cu adatoms on thin films described in details in ref.[29]. By changing the type of the thin insulating film (from purely ionic compound as NaF to covalent compound ScN) one can tune the charge state of metal adatom from nearly $Cu^{1-}$ (in the case of NaF film) to $Cu^0$ (in the case of ScN film). The results are summarized in Table 3. The inspection of the electronic structure of supported Cu atoms (Fig. 4, left) clearly shows tremendous influence of the underlying substrate. This can be attributed to the change of the bonding between the Cu adatom and the substrate from predominantly ionic to predominantly covalent. As expected, the reactivity of Cu atoms is also greatly influenced, which is clear from the calculated CO and H adsorption energies. It is also significantly altered compared to the reactivity of isolated Cu (Table 3), like in the case of considered metals on MgO(001) (Table 1). We see that the CO adsorption energies span in the 2 eV wide energy window, while H adsorption energy ranges between –2.20 and –3.36 eV (Table 3). Moreover, there is no clear connection between the d-band center position and the CO/H adsorption energy – the most reactive Cu adatom is the one supported by ScN@Mo(001) which actually has the d-band located at the lowest energies (Fig. 4). However, the correlation between the charge state of Cu adatom and the adsorption energies holds for both CO and H adsorption (Fig. 4).

Table 3. Cu binding energies to XY@Mo(001) substrate, charge state of Cu adatoms and CO and H adsorption energies on supported Cu adatoms.

| substrate | Cu ads. site[a] | $E_b$(Cu) / eV[a] | $q$(M) / |e|[a] | $E_{ads}$(CO) / eV | $E_{ads}$(H) / eV |
|---|---|---|---|---|---|
| **NaF@Mo(001)** | F on-top | –1.88 | –0.76 | –0.63 | –2.20 |
| **MgO@Mo(001)** | hollow | –1.47 | –0.67 | –0.31 | –2.30 |
| **ScN@Mo(001)** | hollow | –0.91 | –0.37 | –2.00 | –3.08 |
| **ScN@Mo(001)** | N on-top | –1.50 | –0.03 | –2.26 | –3.36 |
| **isolated Cu** | | | 0.00 | –0.78 | –2.88 |

[a]ref.[29]

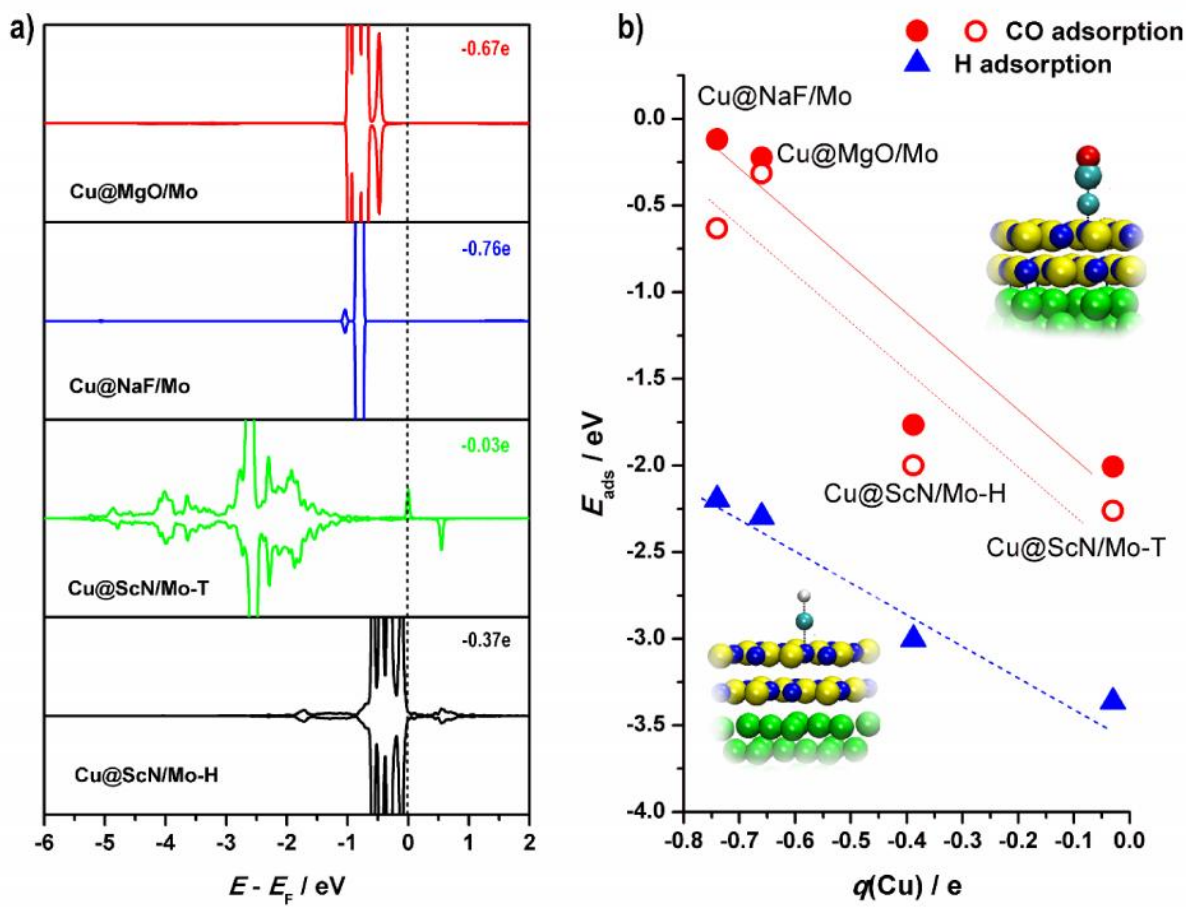

**Figure 4.** (a) Projected densities of states of Cu adatoms adsorbed on Mo-supported thin films of MgO, NaF and ScN (T – N top site, H – hollow site at ScN thin film). Vertical line indicates the Fermi level, while the numbers indicated in the top right corner of each plot give the partial charge of Cu adatom; (b) Correlation between the charge state of supported Cu atom (before atom/molecule adsorption) and the adsorption energy of H (triangles) and CO molecule (full circle – constrained to $C_{nv}$ symmetry, opened circles – non-constrained). The insets represent example CO (top right) and H (bottom left) adsorption structures.

Again, the interaction between the CO molecule and Cu adatom can be rationalized in terms of donation-backdonation mechanism. The more negative charge is located at Cu adatom, the weaker bonding of CO is going to be due to more backdonation to 2π* orbitals of CO. As the charge of adatom decays from –0.76 (NaF@Mo(001) substrate) to –0.67 (MgO@Mo(001) substrate) and –0.03 (ScN@Mo(001) substrate) the net negative charge of adsorbed CO molecule goes from –0.24, to –0.15 and finally –0.10|e|. The interaction between the d-states of Cu adatoms and CO states is clearly seen in Fig. 5. In contrast, $H_{ads}$ interacts with the s-states of Cu (Fig. 5) which also offers an explanation for the observed behavior. Namely, as the s-states of Cu adatom start to fill due to the charge injected from the substrate, Cu adatom is more saturated and tends to bond weaker through its s orbital.

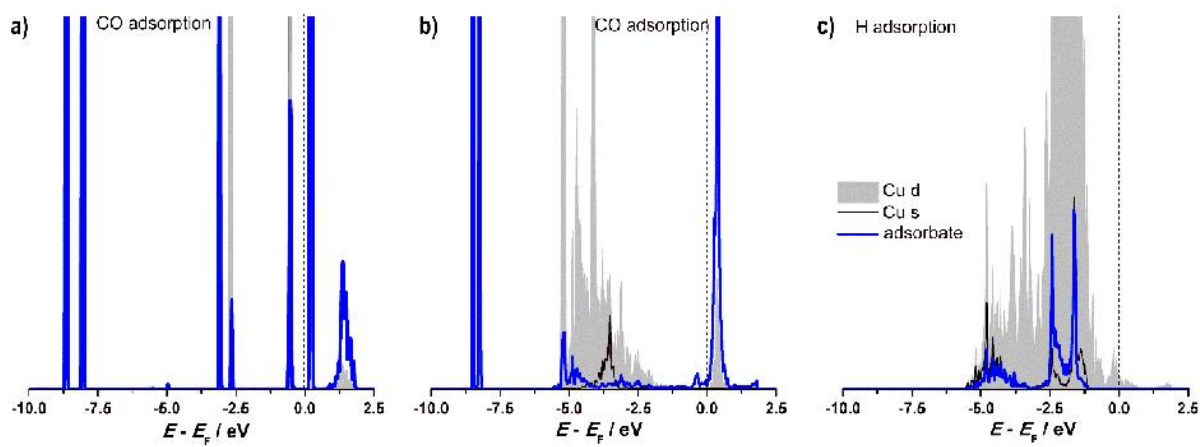

**Figure 5.** PDOS of CO molecule and Cu atom for the case of Cu supported by (a) NaF@Mo(001) and (b) ScN@Mo(001) (Cu adsorbed at N-top site); (c) PDOS of H and Cu atoms for the case of H adsorption on Cu/ScN@Mo(001).

When comparing the electronic structures of the Cu adatom on ScN@Mo(001) with the other cases (Fig. 4), we see a rather wide d-band, which is a consequence of the covalent interaction with underlying ScN. Moreover, these adatoms are very reactive compared to the other cases. Cu supported by ScN@Mo(001) binds H similarly as Cu(111) surface,[62] while it binds CO much stronger than Cu(111).[63] We consider that this is due to the differences in the electronic structure of supported Cu and the Cu(111) surface. It is known that the electronic structure of Cu(111) is characterized by a wide half-filled s-band and narrow, completely filled d-band. Due to the narrow d-band it binds CO very weakly, but the interaction between the wide s-band and the s-states of H makes the interaction stronger. In contrast, a wide d-band of Cu adatom on ScN@Mo(001) makes the interaction with CO strong as well (Figs. 4 and 5). In order to further investigate the reactivity of adatoms covalently bonded to the substrate, we chose graphene-supported Pt atoms, the system which was considered as SAC from both theoretical[43] and experimental[13] viewpoints.

### 3. CO adsorption on graphene and rGO-supported Pt monomers

As a model of the support we use previously described epoxy- and hydroxyl-graphene models.[44,64] As shown before[64] the reactivity of graphene basal plane can be significantly altered by attaching different oxygen functional groups on its surface. This is clearly reflected in enhanced bonding of Pt to oxidized graphene surfaces[64] (see Table 4). While Pt is adsorbed on pristine graphene at the bridge site, this is not a general case for the oxidized surfaces. The introduction of the oxygen surface groups results in strong bonding of Pt adatoms at top sites of C atoms in the vicinity of oxygen-functional groups (Table 4). Depending on the specific adsorption site the charge transferred to Pt adatom can vary but, in general, less charge is transferred to Pt when compared to the case of MgO(001) support (Section 3.1). The position of the d-band center is governed by the interaction with the sp-states of the carbon atom (see Fig. 6) and is typically much higher in energy than the d-band center of Pt.[65] The bonding of CO to single Pt adatoms is much stronger than bonding on Pt(111) surface (−1.48 eV).[58]

Table 4. Adsorption of Pt on pristine and oxidized graphene, charge states and d-band centers of Pt adatoms, and CO adsorption energies on Pt adatoms. The data for the preferential adsorption sites on given surfaces are marked by bold letters.

| surface model | $E_b$(Pt) / eV | Pt ads. Site | $q$(Pt) / e | $E_{d-band}$(Pt) / eV | $E_{ads}$(CO) / eV |
|---|---|---|---|---|---|
| pristine graphene | −1.56 | Bridge | −0.03 | −1.60 | −2.97 |
| epoxy-graphene-1 | **−2.18** | **Top** | **−0.10** | **−1.57** | **−2.59** |
|  | −1.85 | Bridge | −0.11 | −1.72 | −3.00 |
| epoxy-graphene-2 | **−2.82** | **Top** | **−0.43** | **−2.40** | **−2.03** |
|  | −1.71 | Bridge | −0.14 | −1.44 | −2.82 |
|  | −2.21 | Bridge | −0.11 | −1.60 | −2.56 |
| hydroxyl-graphene | **−2.40** | **Bridge** | **−0.05** | **−1.80** | **−2.60** |
|  | −2.03 | Bridge | −0.06 | −1.68 | −2.77 |
|  | −1.73 | Bridge | −0.17 | −1.73 | −3.25 |

In this case we have not observed any correlation with the d-band center position of Pt adatoms. No connection between the charge transferred to Pt adatom and CO adsorption energy is seen, either. In comparison, Liu and Huang have investigated $O_2$ adsorption on Pt single atoms adsorbed on graphene single vacancy and $C_2N$, $CN_2$ and $N_3$ moieties incorporated into the graphene lattice.[66] They have reported Pt d-band centers as low as −4.93 eV for the case of Pt at graphene single vacancy, while good correlation between $O_2$ adsorption energies and d-band centers has been observed. Also, authors observed positive charge on Pt atoms and stronger binding of $O_2$ for more positively charged Pt adatoms.[66] However, in our case we could only correlate the CO adsorption energies to Pt binding energies on model surfaces of graphene (Fig. 6). Our results suggest that weaker bonding of a Pt adatom to the surface results in stronger adsorption of CO on a given adatom. As already stated, this is intuitively clear as a stronger bond between Pt and the substrate means that the adatom is more saturated and, in this way, less reactive. Interestingly, the same correlation can be derived from the results in the ref.[66]. However, their $E_{ads}$(CO@Pt) = $f(E_b$(Pt)) line falls far from the one in Fig. 6. The observed difference can be ascribed to the interaction of Pt atoms with N atoms incorporated in the graphene lattice, being considered as anchoring sites for Pt in ref.[66]. Hence, we also note that chemical environment of supported single atoms is important for its reactivity.

In order to analyze the bonding we turn to the inspection of the electronic structure (Fig. 6). We see general features which are in accordance with already discussed donation-backdonation mechanism. The states of Pt (s and d) clearly interact with the electronic states of the CO molecule. The backdonation of charge to the 2 * state of CO molecule gives rise to the elongation of the C-O bond which was found to be around 1.16 Å in all the considered cases. So, the question is why it is not possible to use the same descriptors as before if the mechanism of the interaction is the same? We have inspected the relaxed structures and observed that there is significant relaxation of the oxidized graphene support upon the adsorption of CO on Pt adatoms. As a result, Pt–C bonds are significantly elongated (in some case even up to 10%) and CO@Pt complex moves along the surface, so that Pt adatom is not actually located at the same adsorption site as before CO adsorption. Moreover, the interaction between CO@Pt complex and the oxygen functional groups located on the surface presents an additional factor that affects both the electronic structure of Pt adatoms, its interaction with the support and the CO adsorption energy. In contrast, in ref.[66] Pt single atoms seem to be much more strongly anchored at the defect sites of graphene basal plane with the adsorption energies between −7.25 and −3.02 eV. As a result, the adatoms remain at the same sites upon the adsorption of $O_2$ which could explain why the correlations with $E_{d-band}$ and the charge of adatom are observed. The presented results turn the focus back to the stability issues, outlined in ref.[1], as the main question related to the performance of SACs. However, another dimension can be added – in order to be able to predict the reactivity of SACs, with the aim to design such a catalyst, the stability of single supported adatom under operating conditions must be ensured.

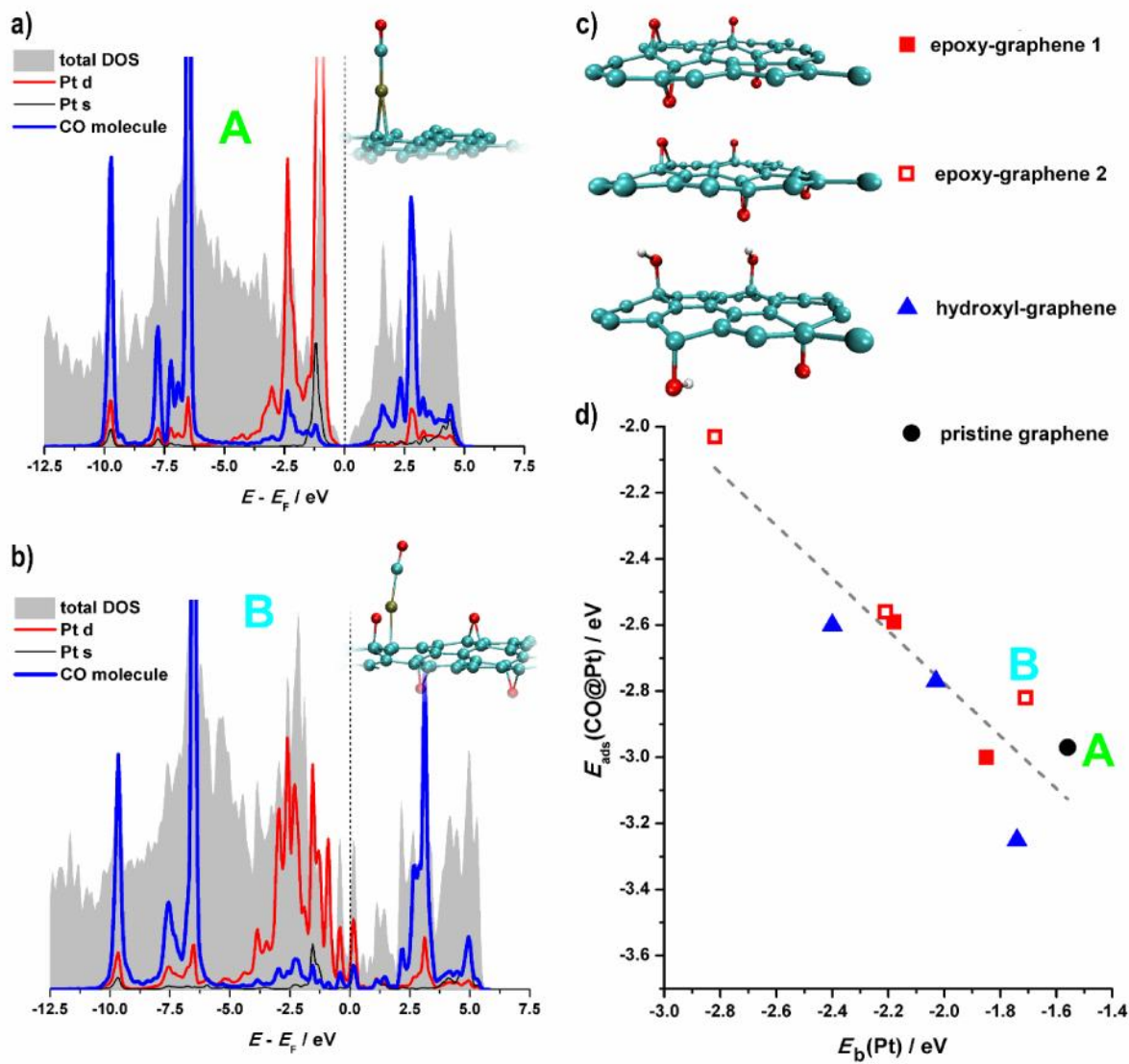

**Figure 6.** PDOS for CO molecule adsorbed on Pt atom supported by pristine graphene (a) and epoxy-graphene (b). Total DOS is also included (vertical line gives the Fermi level). The models of oxidized graphene used in this work (c) and the correlation between CO adsorption energy on supported Pt atom and Pt binding energy (d) are given on the right. The structures whose PDOSes are presented in a) and b) are indicated in d) by (A) and (B).

## 4. Conclusion

We have considered several rather different systems consisted of single supported metal atoms or dimers on different supports as models of single atom catalysts. The systems include Ru, Rh, Pd, Ir and Pt and their dimers on MgO(001), Cu adatoms on different Mo(001)-supported thin films and, finally, single Pt atoms on oxidized graphene surfaces. By testing the reactivity of these single atoms with CO molecule, we observe that single atoms and dimers are more reactive than corresponding metallic surfaces. Moreover, in all the cases donation-backdonation mechanism is found to be operative. However, there is a limited number of potential descriptors which seem to be applicable in the considered cases. For MgO(001)-supported metal dimers, the reactivity of a particular atom can be easily tuned by changing the composition of the dimer, while scaling between the adsorption energies and d-band centers or charge of atoms in dimers was observed. Moreover, if the binding energy of considered atom in the dimer structure is considered as a measure of atom saturation, then less saturated atoms bind CO stronger. The effects of adatom charge on its reactivity was tested for the case of Cu adatom on thin Mo(001)-supported films. We observe that more negatively charged atoms bind CO weaker. The same was the case for the adsorption of H atom, although the mechanism is considered to be different here. For the case of covalently bonded Pt adatoms on oxidized graphene surfaces the d-band center and the adatom charge cannot be connected to CO adsorption energy. In this case we observe a link between Pt binding energy to the support and CO adsorption energy on Pt atoms – more strongly bonded Pt atoms adsorb CO weaker.

Overall, several possible reactivity descriptors of low coordinated/single metal atoms on foreign supports might be outlined. Some of them, like the d-band center or metal atom charge do have physical background and could offer a strategy for a rational design of single atom catalyst. However, this study also shows that the stability of single metal adatoms over foreign substrates is not only of paramount importance for practical use of single atom catalysts but also for the identification of reliable activity descriptors and their use in establishing reactivity trends.

**Conflict of interest**

There is no conflict of interest to declare.

**Acknowledgement**

N.V.S. acknowledges the support provided by Swedish Research Council through the project No. 2014-5993. The computations were performed on resources provided by the Swedish National Infrastructure for Computing (SNIC) at National Supercomputer Centre (NSC) at Linköping University. We also acknowledge the support from Carl Tryggers Foundation for Scientific Research (grant no. 18:177). Authors would like to thank to Dr. Pjotr Žguns for providing the models of layered substrates described in Section 3.2.